\documentclass[twocolumn]{aastex62}

\submitjournal{ApJL}

\shorttitle{whistlers in the solar wind}
\shortauthors{Tong et al.}

\begin{document}

\title{Whistler wave generation by halo electrons in the solar wind}

\correspondingauthor{Yuguang Tong}
\email{ygtong@berkeley.edu}

\author{Yuguang Tong}
\affil{Space Sciences Laboratory, University of California, Berkeley, CA 94720}
\affil{Physics Department, University of California, Berkeley, CA 94720}

\author{Ivan Y. Vasko}
\affil{Space Sciences Laboratory, University of California, Berkeley, CA 94720}

\author{Marc Pulupa}
\affil{Space Sciences Laboratory, University of California, Berkeley, CA 94720}

\author{Forrest S. Mozer}
\affil{Space Sciences Laboratory, University of California, Berkeley, CA 94720}

\author{Stuart D. Bale}
\affil{Space Sciences Laboratory, University of California, Berkeley, CA 94720}
\affil{Physics Department, University of California, Berkeley, CA 94720}

\author{Anton V. Artemyev}
\affil{nstitute of Geophysics and Planetary Sciences, University of California, Los Angeles, USA}

\author{Vladimir Krasnoselskikh}
\affil{University of Orleans, Orlean, France}



\begin{abstract}
We present an analysis of simultaneous particle and field measurements from the ARTEMIS spacecraft which demonstrate that quasi-parallel whistler waves in the solar wind can be generated locally by a bulk flow of halo electrons (whistler heat flux instability). ARTEMIS observes quasi-parallel whistler waves in the frequency range $\sim 0.05 - 0.2 f_{ce}$ simultaneously with electron velocity distribution functions that are a combination of counter-streaming core and halo populations.  A linear stability analysis shows that the plasma is stable when there are no whistler waves, and unstable in the presence of whistler waves. In the latter case, the stability analysis shows that the whistler wave growth time is from a few to ten seconds at frequencies and wavenumbers that match the observations. The observations clearly demonstrate that the temperature anisotropy of halo electrons crucially affects the heat flux instability onset: a slight anisotropy $T_{\parallel}/T_{\perp}>1$ may quench the instability, while a slight anisotropy $T_{\parallel} / T_{\perp}<1$ may significantly increase the growth rate. These results demonstrate that heat flux inhibition is strongly dependent on the microscopic plasma properties.

\end{abstract}



\section{Introduction}

The mechanisms controlling the heat flux in collisionless or weakly-collisional plasmas are of high interest in astrophysics \citep{Cowie77,Pistinner1998a,Roberg-Clark2018a}. In-situ measurements in the solar wind indicate that the heat flux is generally different from the classical Spitzer-H{\"a}rm prediction \citep{Feldman75,Scime94,Bale13} and apparently constrained by a threshold dependent on local plasma parameters \citep{Gary1999b,Gary2000a,Tong18}. Such observations have motivated many studies on the detailed physics of heat flux inhibition in the solar wind. 

In the slow solar wind ($v_{sw}\lesssim 400$ km/s) the electron velocity distribution function can often be approximated by a bi-Maxwellian thermal dense core and a tenuous, suprathermal halo \citep{Feldman75,Maksimovic97}. The heat flux is predominantly parallel to the magnetic field and carried by suprathermal electrons.  Linear stability analysis shows that the  counter-streaming core and halo electrons are capable of driving whistler waves propagating quasi-parallel to the bulk flow of the halo population via the so-called heat flux instability \citep{Gary75,Gary1994a, Gary2000a}. The quasi-linear theory \citep{Gary77,Pistinner1998a} and numerical simulations \citep{Roberg-Clark2018a,komarov_2018, Roberg-Clark:2018b} suggest that the scattering of halo electrons by the whistler waves should suppress the heat flux below some threshold value that is in general agreement with the heat flux constraints observed in the solar wind \citep{Gary1994a, Gary1999b,Tong18}. However, the aforementioned experimental studies did not provide measurements of whistler waves accompanying the electron heat flux measurements, and are therefore insufficient to firmly establish the heat flux inhibition by whistler waves in the solar wind.

It is not until recently that careful studies of whistler waves presumably generated by the heat flux instability in freely expanding solar wind have been reported with measurements on Cluster and ARTEMIS spacecraft. \cite{Lacombe14} reported whistler waves observed along with the heat flux values close to the theoretical threshold given by \cite{Gary1999b}. \cite{Stansby16} presented observations of similar whistler waves on ARTEMIS and determined the dependence of the whistler wave dispersion relation on $\beta_{e}$. However, neither study showed that the whistler waves were indeed generated by the heat flux instability in the local plasma, leaving the possibility that whistler waves were generated in a very different plasma by an alternative mechanism and propagated to the spacecraft location. We note that whistler waves in the solar wind can be associated with shocks and stream interaction regions \citep{LengyelFrey96, Lin1998a,Breneman10,Wilson13}, while we focus on whistler waves in the freely expanding solar wind.  

In this study we analyze simultaneous particle and wave measurements for data intervals presented by \cite{Stansby16} and carry out linear stability analysis on electron velocity distribution functions. We find that the observed whistler waves are indeed generated locally by the heat flux instability on a time scale of a few seconds. In this letter we present one of those events, which also demonstrates crucial features of the heat flux instability. 

\section{Observations}

We consider observations of ARTEMIS \citep{Angelopoulos2011a} on November 9, 2010 for ten minutes around 10:17:00 UT as the spacecraft was in the pristine solar wind about 40 Earth radii upstream of the Earth's bow shock. We use measurements of the following instruments aboard ARTEMIS: the magnetic fields with 3 second resolution provided by the Flux Gate Magnetometer \citep{Auster08}, the electron velocity distribution function (32 log-spaced energy bins from a few eV up to 25 keV and 88 angular bins) and particle moments (density, bulk velocity and temperatures) with 3 second time resolution provided by the Electrostatic Analyzer \citep{McFadden08}, measurements of three magnetic and electric field components at 128 Hz sampling rate provided by the Search Coil Magnetometer \citep{LeContel08} and Electric Field Instrument \citep{Bonnell08}. 

\begin{figure*}
    \centering
    \includegraphics[width=1.0\linewidth]{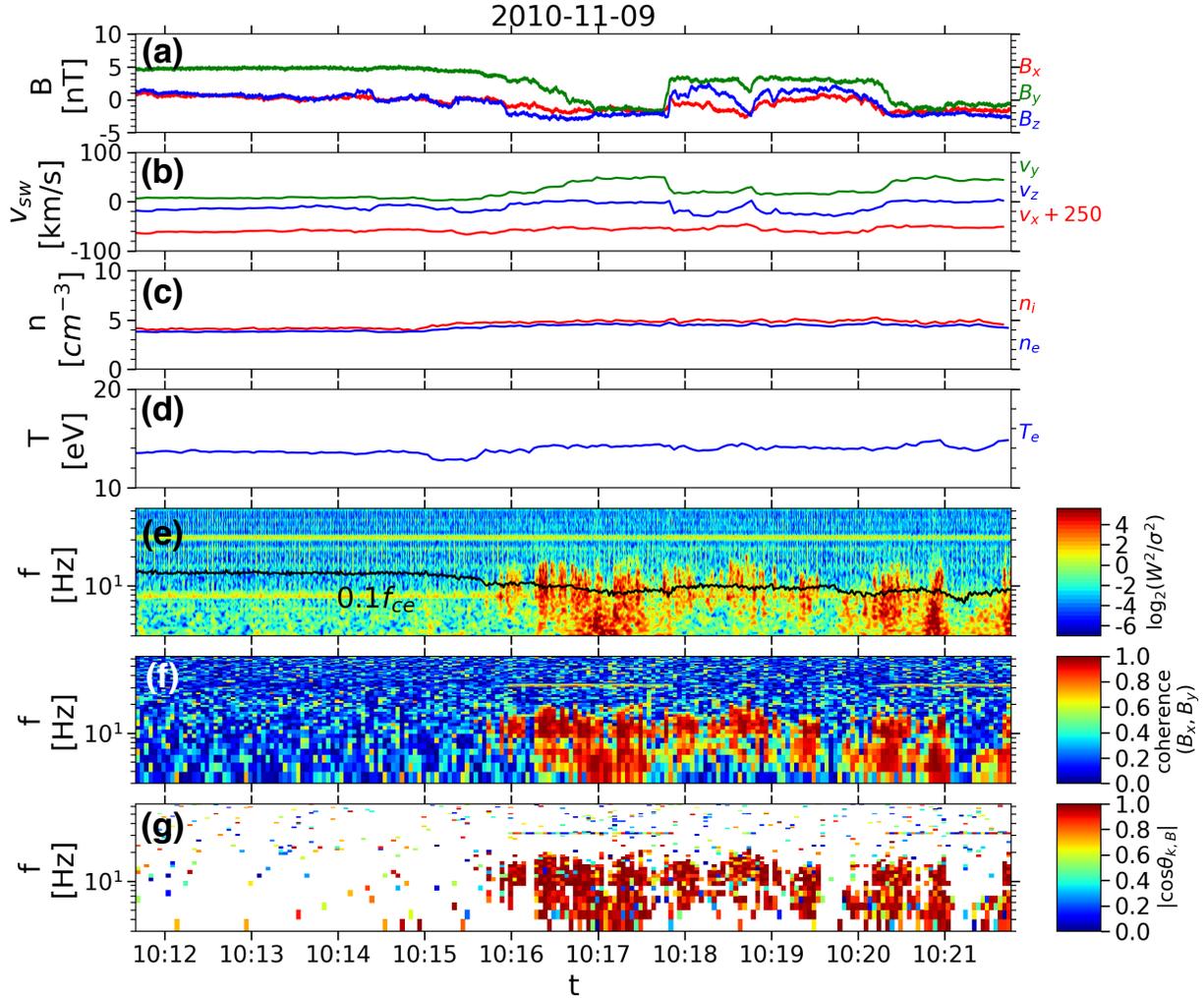}
    \caption{ARTEMIS observations in the pristine solar wind on November 9, 2010 about 40 Earth radii upstream of the Earth's bow shock: (a) quasi-static magnetic field; (b) ion bulk velocity in the GSM coordinate system; (c,d) electron and ion densities and temperatures; (e) wavelet power spectrum of one of the magnetic field components perpendicular to the quasi-static magnetic field;  we use a Morlet wavelet with center frequency $\omega_0=32$ as the mother wavelet and normalize the wavelet power ($W^2$) by the white noise power ($\sigma^2$); (f) the coherence coefficient between magnetic field components $B_x$ and $B_y$ perpendicular to the quasi-static magnetic field; (g) $|\cos\theta_{\bf k B}|$ indicating obliqueness of the whistler waves (${\bf k}$ and ${\bf B}$ are the wave vector and the quasi-static magnetic field). In panel (g) domains with coherence smaller than 0.6 have been masked out for clarity. 2D maps (e)-(g) are computed using the magnetic field measured at 128 Hz sampling rate. \label{fig1}}
\end{figure*}

Figure \ref{fig1} shows that the solar wind was streaming at about $v_{sw}\sim 320$ km/s, the quasi-static magnetic field was gradually decreasing from $B_0\sim 5$ nT to 3 nT, the plasma density was $n_0\sim 5$ cm$^{-3}$, and the electron temperature was $T_{e}\sim 15$ eV. The ion temperature was $T_i\sim 5$ eV (from the OMNI dataset and not shown). The electron cyclotron frequency $f_{ce}$ was varying from 150 to 90 Hz, the Alfv{\'e}n speed $v_{A}=B_0/(4\pi n_0 m_i)^{1/2}$ from 90 to 30 km/s, while $\beta_{i,e}=8\pi n_0 T_{i,e}/B_0^2\sim 0.5-2$. Over the ten minute interval, continuous electric and magnetic field measurements at 128 samples per second were available. Panel (e) presents the wavelet power spectrum of one of the magnetic field components perpendicular to the quasi-static magnetic field. The enhancement of spectral power density from a few Hz up to about 0.2 $f_{ce}$ corresponds to whistler waves \citep{Stansby16}. Since the power spectra above 64 Hz cannot be obtained from the search coil magnetic field time series,  we also checked the on board FFT power spectra of search coil magnetic fields (not shown) covering 8 Hz-4 kHz and verified that there was no significant power between 64 Hz and $f_{ce}$.  Panel (f) presents the spectral coherence between the two magnetic field components perpendicular to the quasi-static magnetic field and indicates a high coherence of the whistler waves. We carry out a spectral polarization analysis following \citet{Santolik03} to determine the obliqueness of whistler waves to the quasi-static magnetic field. Panel (g) presents the cosine of the propagation angle and confirms that whistler waves propagate almost parallel or anti-parallel to the quasi-static magnetic field in accordance with the conclusions of \cite{Stansby16}. \added{The amplitude of whistler waves ranges from 0.05-2 nT, and is small compared to $B_0$.}

\begin{figure*}
    \centering
    \includegraphics[width=0.53\linewidth]{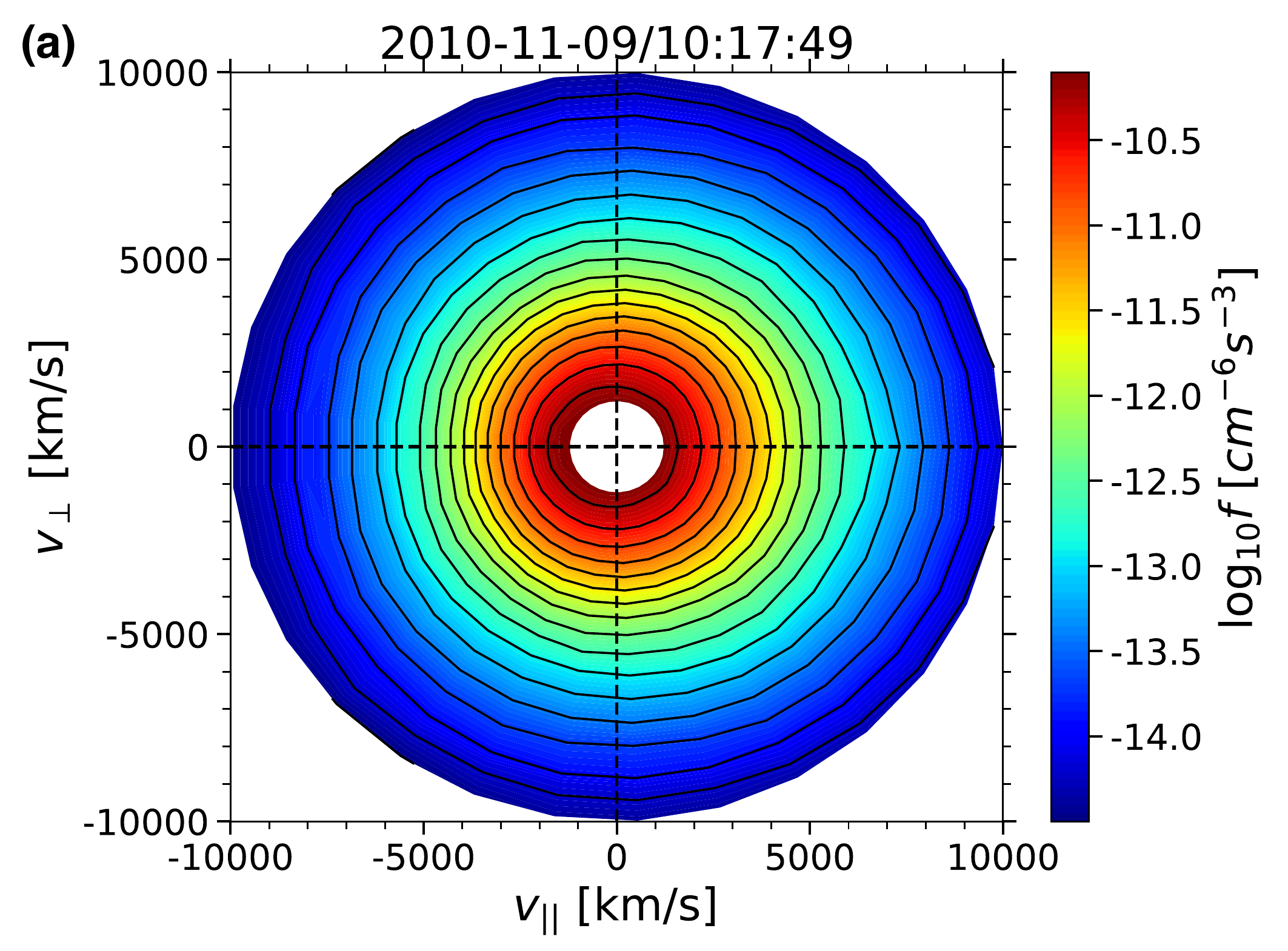}
    \includegraphics[width=0.4\linewidth]{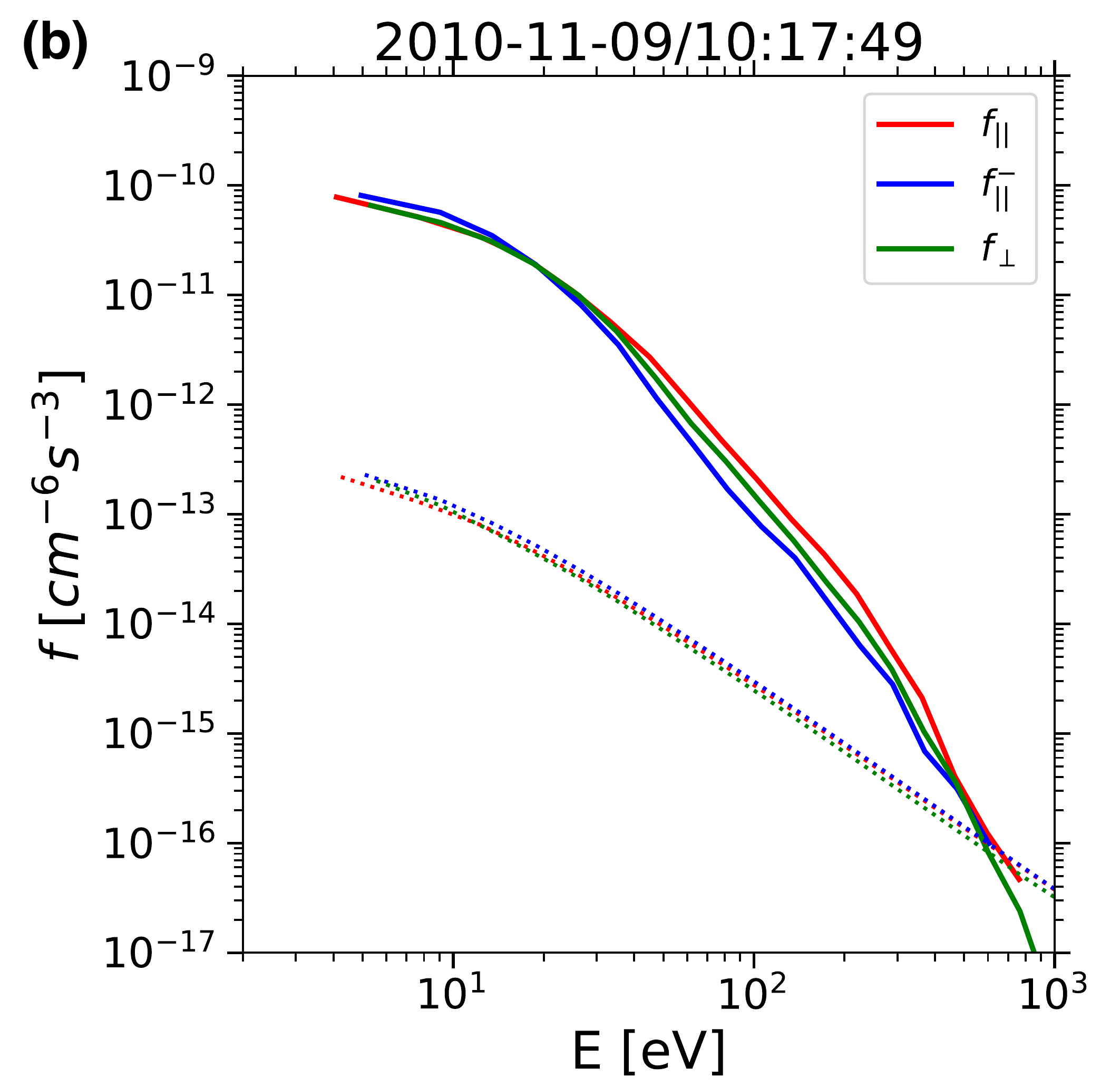}
    \caption{Example of an electron VDF that has been transformed into the solar wind frame and calibrated for the spacecraft potential: (a) gyrophase averaged $f(v_{||},v_{\perp})$, where $v_{||}$ and $v_{\perp}$ are parallel and perpendicular to the magnetic field; (b) VDF cuts plotted vs. electron energy and corresponding to electrons streaming parallel ($f_{||}=f(v_{\parallel}>0,v_{\perp}=0)$), perpendicular ($f_{\perp}=f(v_{\parallel}=0,v_{\perp})$) and anti-parallel ($f_{||}^-=f(v_{\parallel}<0,v_{\perp}=0)$) to the quasi-static magnetic field.\label{fig3}}
\end{figure*}

Figure \ref{fig3} presents an example of the processed electron velocity distribution function (VDF). The raw electron VDF measured around 10:17:49 UT is corrected for the effect of spacecraft potential and transformed from the spacecraft frame into the solar wind frame using the ion bulk velocity measurements. Panel (a) shows the processed VDF $f(v_\parallel, v_\perp)$ averaged over the gyrophase, where $v_{||}$ and $v_{\perp}$ correspond to velocities parallel and perpendicular to the background magnetic field. The VDF is asymmetric in the direction parallel to the magnetic field with opposite asymmetries below and above a few thousand km/s, indicating counter-streaming of cold and hot electrons. Panel (b) shows VDF cuts $f_{||}$, $f_{\perp}$ and $f_{||}^-$ corresponding to electrons streaming parallel (pitch angles $\alpha\sim 0^{\circ}$), perpendicular ($\alpha\sim 90^\circ$) and anti-parallel ($\alpha\sim 180^\circ$) to the quasi-static magnetic field. Below $\sim$30 eV, $f_{||}^{-}>f_{||}$, consistent with core electrons streaming anti-parallel to the magnetic field.  At higher energies,  $f_{||}>f_{||}^{-}$ shows that the hotter electrons are streaming in the opposite direction.

The counter-streaming cold and hot electrons persist through the whole ten minutes in Figure \ref{fig1}. Is this plasma indeed capable of generating the observed whistler waves? How fast is the instability? What controls the absence of whistler waves before 10:16:00 UT and their later appearance? To address these questions we fit the processed electron VDFs and carry out a linear kinetic stability analysis using the previously developed numerical code \citep{Tong15}.

\section{Analysis}

During this slow solar wind interval, the electron VDFs are well described by a combination of core and halo populations $f=f_c+f_h$. The core and halo are modelled respectively with drifting bi-Maxwellian and bi-kappa distributions
\begin{eqnarray*}
    f_c&=&A_c\exp\left[-\frac{m_e\left(v_{||}-v_{0c}\right)^2}{2T_{||c}} - \frac{m_e v_{\perp}^2}{2T_{\perp c}}\right],\\
    f_h&=&A_h B_{\kappa}\left[1 + \frac{m_e(v_{||}-v_{0h})^2}{(2\kappa-3)T_{||h}}+\frac{m_ev_{\perp}^2}{(2\kappa-3)T_{\perp h}}\right]^{-(\kappa+1)},
\end{eqnarray*}
where $A_{s}=n_s (m_e/2\pi T_{\perp s}^{2/3}T_{||s}^{1/3})^{3/2}$, $B_{\kappa}=\Gamma(\kappa+1)/(\kappa-3/2)^{3/2}\Gamma(\kappa-1/2)$ and $n_{s}$, $v_{0s}$, $T_{\perp s}$, $T_{||s}$ are densities, bulk velocities and temperatures (parallel and perpendicular to the quasi-static magnetic field ${\bf B}_0$) of the core and halo populations ($s=c,h$). These parameters are estimated by fitting the model to VDF cuts $f_{||}$, $f_{\perp}$ and $f_{||}^-$ using the standard $\chi^2$ minimization method. Following \cite{Feldman75} the electron current in the solar wind frame is kept zero by restricting the parameters to $n_cv_{0c}+n_hv_{0h}=0$.

\begin{figure*}
  \includegraphics[width=0.45\linewidth]{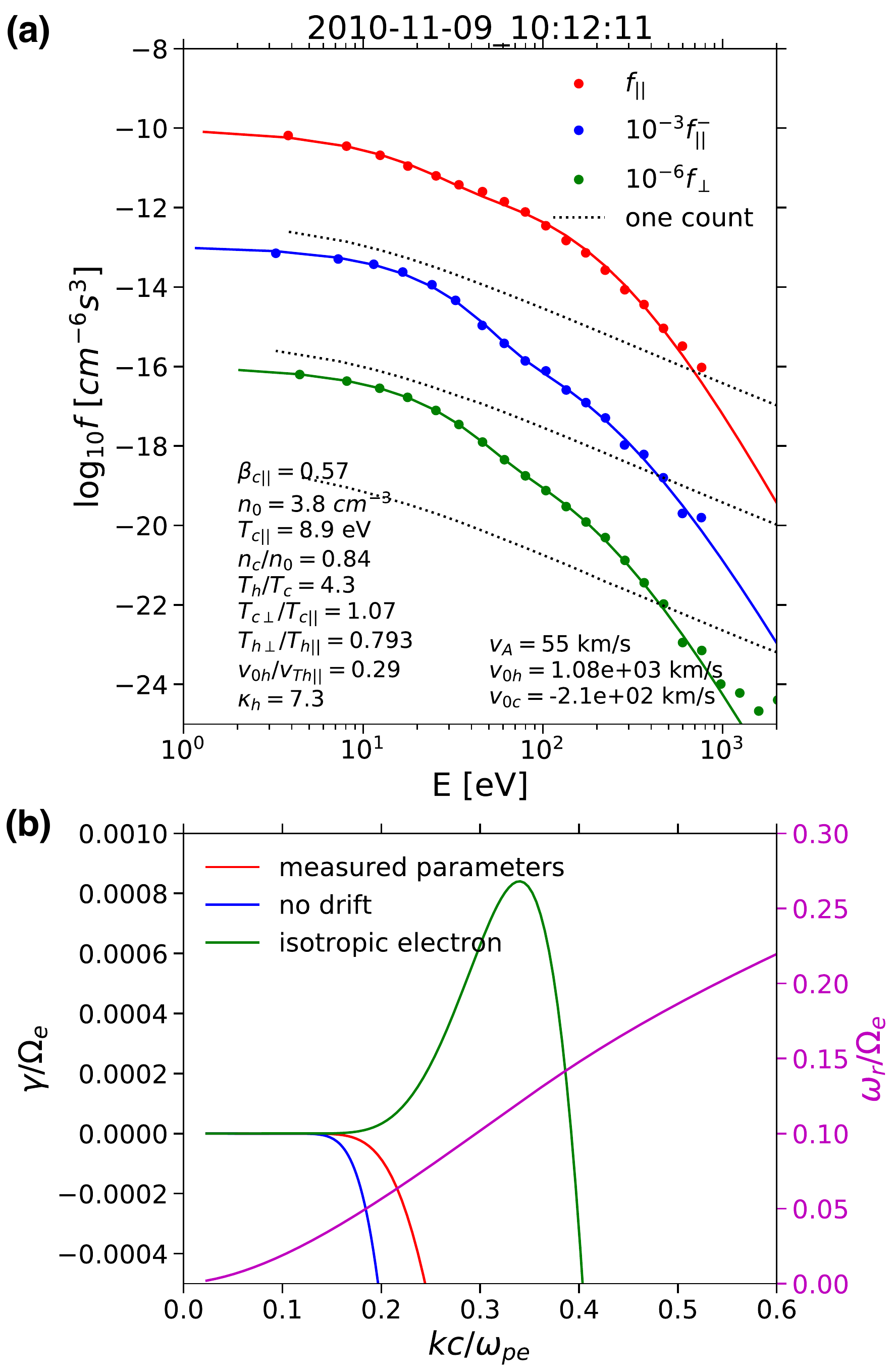}
  \includegraphics[width=0.45\linewidth]{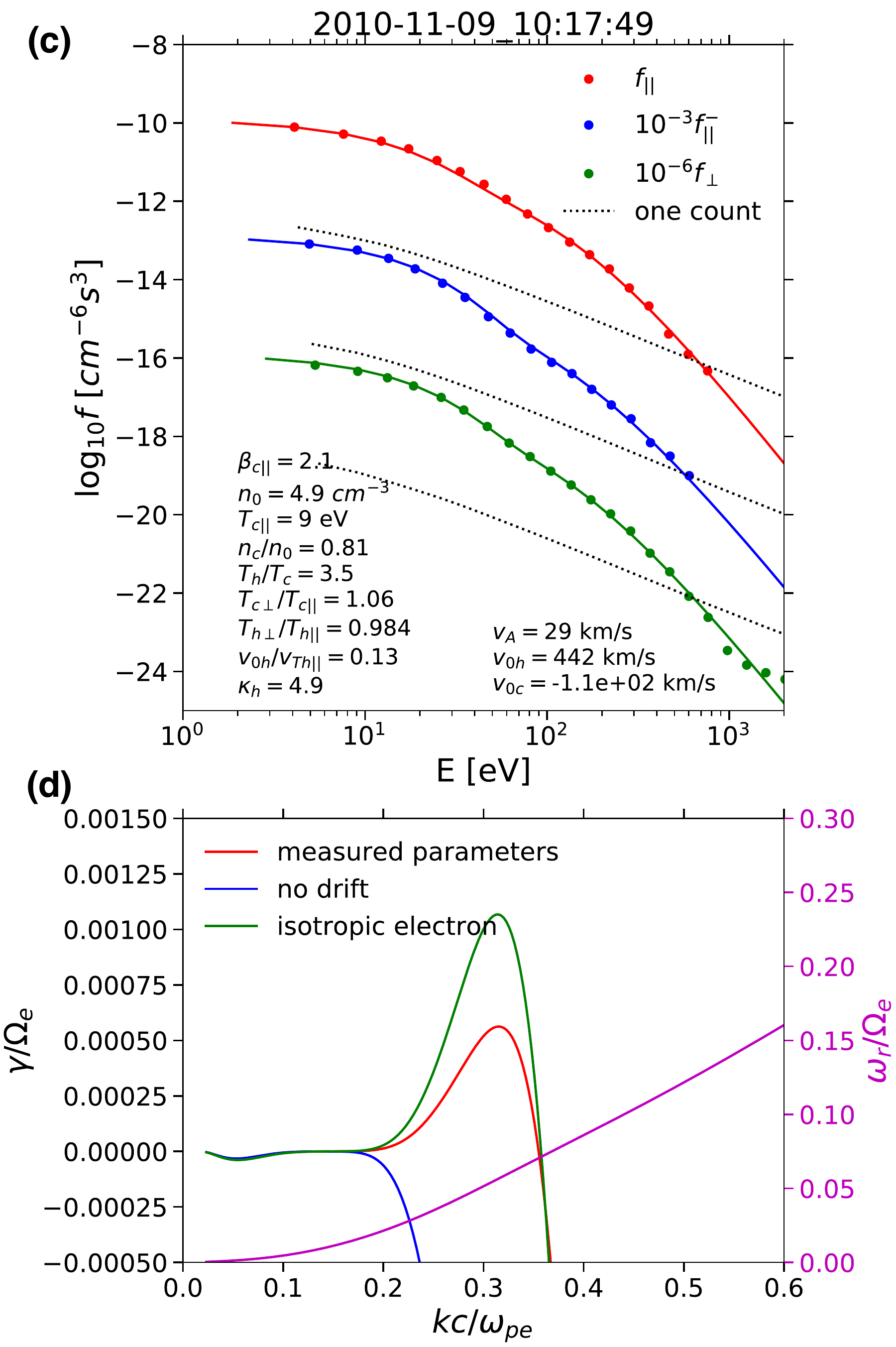}
  \caption{Illustration of the fitting procedure and linear stability analysis of VDFs associated with negligible and noticeable whistler wave activity observed around 10:12:11 and 10:17:49 UT: (a,c) the VDF cuts $f_{||}$, $f_{\perp}$ and $f_{||}^-$ corresponding to electrons with pitch angles around $0^\circ$, $90^\circ$ and $180^\circ$ are shown with dots; the VDF cuts are shifted vertically with respect to each other for visual clarity; only VDF values above one count level (dashed curves) have been used in the fitting procedure; the model fits are presented with solid curves and the fitting parameters are indicated in the panels; (b,d) the growth rate and dispersion curves of parallel propagating whistler waves; the growth rate computations are carried out for (red) the measured electron VDFs and for the measured VDF with either (blue) core and halo bulk velocities set to zero or (green) temperature-isotropic core and halo.}
  \label{fig:example_fits}
\end{figure*}

Figures \ref{fig:example_fits} (a) and (c) illustrate the fitting procedure, using an electron VDF measured in absence of whistler wave activity at 10:12:11 UT and another VDF in presence of whistler waves at 10:17:49 UT. Panels (a) and (c) present the VDF cuts, the model fits and the best fit parameters. Only data points above the one count level have been used in the fitting procedure. Core electrons make up about 80-85\% of the total electron density, the bulk velocity is 100-200 km/s (anti-parallel to ${\bf B}_0$ in the solar wind frame), or about four times larger than the local Alfv{\'e}n speed, the temperature is around 9 eV, and the parallel and perpendicular temperatures are slightly different, $T_{\perp c}/T_{|| c}\sim 1.06$. The halo bulk velocity is about 500-1000 km/s (parallel to ${\bf B}_0$) and the temperature is about 30 eV. The halo population is rather anisotropic in (a) with $T_{\perp h}/T_{\parallel,h}\sim 0.8$ and essentially isotropic in (c) with $T_{\perp h}/T_{\parallel,h}\sim 1.0$.

We address the whistler wave generation by carrying out a linear stability analysis. In the computations we use the model electron VDF with the best fit parameters and isotropic Maxwellian protons with a temperature of 5 eV. The precise shape of the ion distribution function is not critical, because thermal ions do not interact resonantly with the observed whistler waves. We have restricted computations to parallel propagating whistler waves, because counter-streaming core and halo electrons with parameters realistic to the solar wind are known to generate whistler waves propagating only quasi-parallel to the bulk flow of the halo population \citep{Gary75,Gary1994a}, which is parallel to ${\bf B}_0$ in our case.

Figures \ref{fig:example_fits} (b) and (d) present growth rates and dispersion curves of parallel propagating whistler waves computed for electron VDFs in (a) and (c). In agreement with observations we find whistler waves to be stable for VDF (a), but unstable for VDF (c). In the latter case the linear stability analysis predicts the fastest growing whistler waves at the frequency of 0.05 $f_{ce}$. Although it is in general agreement with the whistler wave spectrum in Figure \ref{fig1}e, a careful comparison requires Doppler shifting the plasma frame frequency of 0.05 $f_{ce}$ into the spacecraft frame (see below). Panel (d) shows that the maximum growth rate is about 10$^{-3}f_{ce}$ or $0.5$ s$^{-1}$ in physical units, which corresponds to an e-folding time of about a second.  During this time, whistler waves can only propagate a few hundred kilometers, because the phase velocity of the whistler waves is about $c (f/f_{ce})^{1/2}f_{ce}/f_{pe}\sim $500 km/s, where $f$ and $f_{pe}$ are whistler and plasma frequencies (see also \cite{Stansby16}). This indicates that the observed whistler waves were likely generated locally. 

In order to uniquely identify the free energy source driving the whistler waves, we computed growth rates for electron VDFs (a) and (c), but with either 1) core and halo bulk velocities set to zero or 2) temperature-isotropic core and halo. Panels (b) and (d) show that the electron VDFs with zero bulk velocities (blue curves) are stable and can not generate whistler waves. The free energy driving the observed whistler waves is hence provided by the bulk motions of the core and halo or, in other words, by the electron heat flux. The assumption of isotropic core and halo makes VDF (a) unstable, demonstrating thereby that $T_{||h}/T_{\perp h}>1$ acts to suppress and possibly quench the instability \citep{Gary77}.

\begin{figure*}
    \centering
    \includegraphics[width=\linewidth]{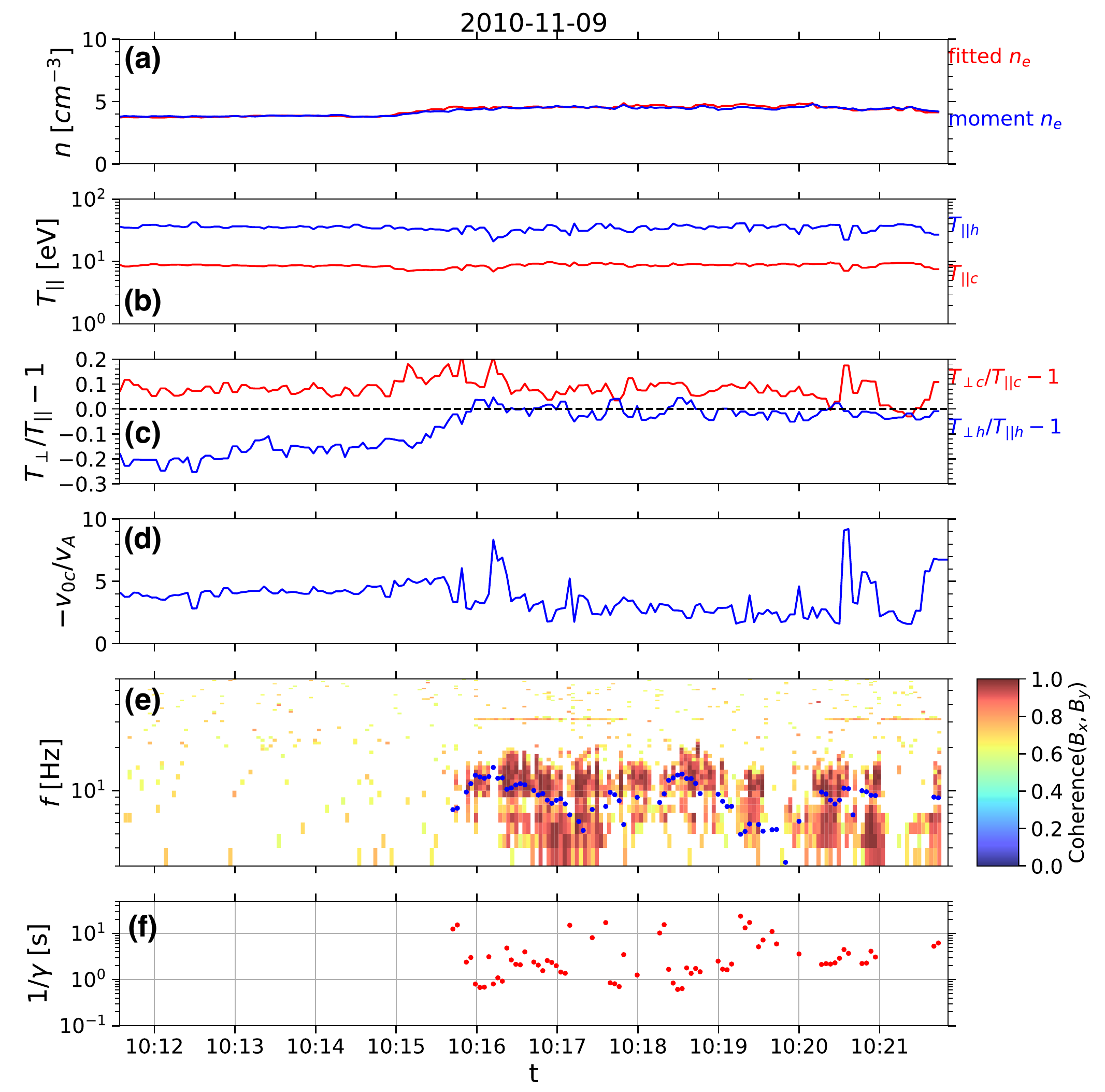}
    \caption{The results of the fitting of 183 electron VDFs: (a) the total electron densities from the fitting and the electron density calibrated on the ground; (b,c) parallel temperatures and temperature anisotropies of the core and halo population; (d) the bulk velocity of core population $v_{0c}$ with respect to the local Alfven speed $v_{A}$. Panel (e) repeats Figure \ref{fig1}g that shows the coherence between the two magnetic field components perpendicular to the quasi-static magnetic field (domains with coherence smaller than 0.6 have been masked out for visual clarity). The spacecraft frame frequency of the fastest growing whistler mode is indicated in panel (e) with dots. Panel (f) presents the e-folding time (inverse of the growth rate) of the fastest growing whistler wave. The absence of dots in some intervals implies that the plasma was stable.}
    \label{fig:stability_analysis}
\end{figure*}

Figures \ref{fig:stability_analysis} (a)-(d) summarize the results of the fitting of all 183 electron VDFs available over the ten-minute interval. Panel (a) demonstrates that the total electron density derived from the fitting matches (within 5\%) the calibrated electron moment densities shown previously in Figure \ref{fig1} (a). Panel (b) shows that the core and halo parallel temperatures are steady. Panel (c) demonstrates that the core temperature anisotropy $T_{\perp c}/T_{||c}$ is steady and around 1.1, while the halo is temperature-anisotropic with $T_{\perp h}/T_{||h}\sim 0.8$ before 10:15:00 UT, gradually becoming isotropic at 10:16:00 UT, and remaining nearly isotropic until the end of the interval. Panel (d) shows that the bulk velocity of the core population varies between 2 and 7$v_{A}$. We perform the linear stability analysis on every electron VDF and determine the growth rate $\gamma_{m}$, frequency $f_{m}$ and wavenumber $k_{m}$ of the fastest growing whistler wave. In the spacecraft frame the whistler wave will be observed at a Doppler-shifted frequency $f_{m} + {\bf k}_{m}\mathbf{\bf v}_{sw}$, where ${\bf k}_{m}$ is parallel to the quasi-static magnetic field ${\bf B}_0$.

Panel (e) demonstrates that the Doppler-shifted frequency of the fastest growing whistler wave indeed traces the observed whistler waves. There are no whistler waves before about 10:16:00 UT, while the plasma is stable. Whistler waves suddenly appear around 10:16:00 UT, when the plasma becomes unstable. Around 10:21:00 UT the plasma is stable for a short time interval, and the coherent whistler waves disappear over this interval. The strong correlation between whistler waves and the local plasma stability/instability indicates that the whistler waves are indeed generated locally. Panel (f) strengthens this conclusion by demonstrating that the e-folding time $\gamma_{m}^{-1}$ of the fastest growing whistler wave is from 1 to 10 seconds.

\begin{figure*}
    \centering
    \includegraphics[width=\linewidth]{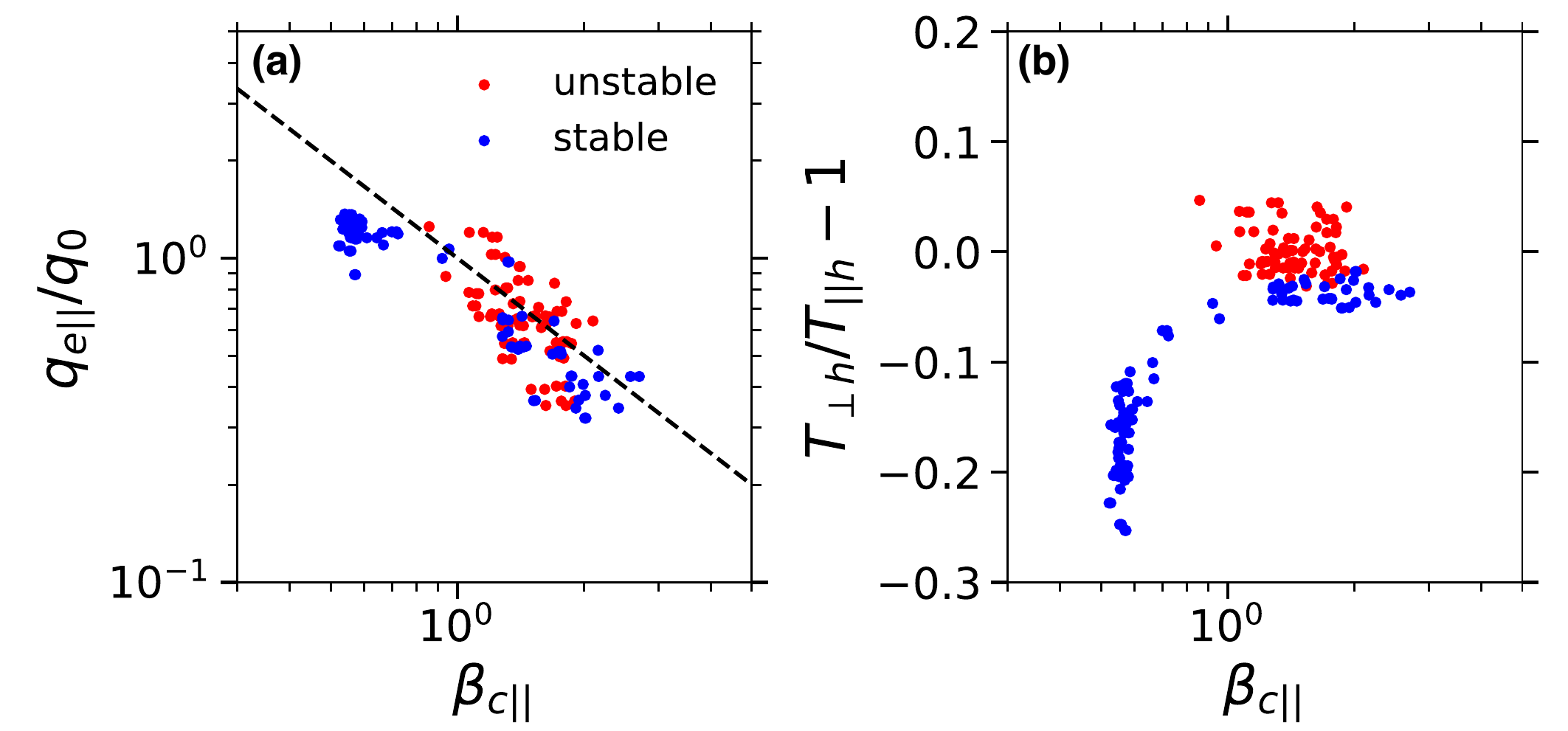}
    \caption{The demonstration of the crucial effect of the halo temperature anisotropy on the whistler heat flux instability. Panel (a) presents the electron heat flux $q_{e}$ normalized to the free-streaming heat flux $q_0=1.5 n_eT_e(T_e/m_e)^{3/2}$ versus core electron beta parameter $\beta_{c||}$  computed for all 183 VDFs available over the ten-minute interval. $q_e/q_0=1/\beta_{c||}$ is plotted in dashed line for reference. Panel (b) presents the temperature anisotropy of the halo population versus $\beta_{c||}$. Unstable (stable) VDFs are labeled with red (blue) dots.}
    \label{fig:heatflux}
\end{figure*}

The abrupt transition from stable to unstable plasma around 10:16:00 UT coincides with the halo population becoming more isotropic. As we demonstrated in Figure \ref{fig:example_fits}, the reason is that the halo temperature anisotropy quenches the whistler heat flux instability. The crucial role of the temperature anisotropy is further demonstrated in Figure \ref{fig:heatflux}. Panel (a) presents the electron heat flux $q_e$ normalized to the free streaming heat flux $q_0=1.5 n_e T_e (T_e/m_e)^{1/2}$ versus $\beta_{c||}=8\pi n_cT_{||c}/B_0^2$. At any given $\beta_{c||}$ the heat flux is clearly below a threshold given by $q_{e}/q_0\sim 1/\beta_{c||}$, that is similar to the marginally stable values in literature \citep{Gary1999a, Pistinner1998a, Roberg-Clark2018a, komarov_2018, Roberg-Clark:2018b}. However, at a given $q_{e}/q_0$ both stable and unstable VDFs are observed, indicating thereby that some other parameter controls the onset of the whistler wave generation. Panel (b) shows that the halo temperature anisotropy separates stable and unstable VDFs with a similar heat flux value. This re-emphasizes the crucial effect of the halo temperature anisotropy on the heat flux constraints in the solar wind.

\section{Discussion and Conclusion}

In-situ observations indicated that whistler waves generated by the heat flux instability highly likely constrain the heat flux in the solar wind \citep{Feldman75,Gary1999b,Tong18}. However, there have been no previous analyses that would prove that whistler waves in the solar wind are actually produced locally by the whistler heat flux instability. In this letter we have presented a careful analysis of simultaneous particle and wave measurements for one of the time intervals in \cite{Stansby16}. We have performed similar analysis for other \cite{Stansby16} time intervals and confirmed that whistler waves are generated locally by the heat flux instability in those intervals as well. The presented event has shown that the e-folding growth time of whistler waves can be as short as one second and clearly demonstrated the crucial effect of the halo temperature anisotropy $T_{\perp h}/T_{||h}<1$. In some of the \cite{Stansby16} events the halo population has $T_{\perp h}/T_{||h}>1$. The linear stability analysis has shown that even a slight $T_{\perp h}/T_{||h}>1$ significantly enhances the growth rate of the heat flux instability, but we stress that the observed temperature anisotropies are insufficient to drive whistler waves  purely via the temperature-anisotropy (without core and halo bulk motion) \citep{Sagdeev60,Kennel66}. 
\added{Other parameters such as plasma beta are also crucial to the onset of whistler waves. The work to find the most critical parameter to whistler heat flux instabilities by statistical studies is under active investigation at this moment and beyond the scope of this letter.} 
The presented analysis indicates that the sporadic occurrence of whistler waves in the solar wind pointed out by \cite{Lacombe14} may be due to an interplay between the electron heat flux and the halo temperature anisotropy that may easily quench or enhance the instability. Future statistical studies should carefully address the halo temperature anisotropy in any analysis of the source of whistler waves in the solar wind.

Up to this point we have been focused on the electron heat flux constrained by wave-particle interactions. In fact, Coulomb electron-electron collisions can also affect solar wind electrons and constrain the electron heat flux \citep{Salem2003, Bale13, Pulupa:2014, Landi:2014a}. The Knudsen number for the observed solar wind $K_n\sim 1 - 1.5$ falls into the collisionless regime \citep[c.f. Figure 2 in ][]{Bale13}. Consistently, the observed heat flux is  30-50\% lower than the Spitzer-H{\"arm} prediction. This implies that the observed heat flux constraint and deviation from the Sptizer-H{\"a}rm prediction are due to electron scattering by the whistler waves. 

Finally, the presented whistler waves are observed in the slow solar wind, where the electron VDF is satisfactorily described by counter-streaming core and halo \citep{Feldman75,Maksimovic97}. In the fast solar wind there is an additional anti-sunward propagating strahl population \citep{Pilipp87,Stverak09} that do not directly interact with parallel whistler waves driven by whistler heat flux instabilities. Hence we expect the whistler heat flux instabilities to operate in the fast wind as well.

\acknowledgments
We acknowledge the THEMIS team for the use of data. We thank T. A. Bowen, J. W. Bonnel, J. M. McTiernan and A. Hull for useful discussions. Y. T. and S. D. B. were supported in part by NASA contract NNN06AA01C. I. V. and F. M. were supported by Johns Hopkins University/Applied Physics Lab Contract No. 922613 (Radiation Belt Storm Probes-Electric Fields and Waves).

\end{document}